\documentstyle[aps,preprint,epsf]{revtex}
\newcommand{\be}{\begin{equation}}
\newcommand{\ee}{\end{equation}}

\newcommand{\ve}[1]{\underline{#1}}

\def\pt2{\partial_t^2}

\def\b2{\beta_2}

\title{A Lattice Boltzmann Model of
Binary Fluid Mixtures}
\author{Enzo Orlandini, Michael R. Swift and J. M. Yeomans \\
	Theoretical Physics, 1 Keble Road, Oxford OX1 3NP, U.K.}

\begin{document}

\maketitle
\begin{abstract}
We introduce a lattice Boltzmann model for simulating
an immiscible binary fluid mixture. Our collision rules
are derived from a macroscopic thermodynamic description
of the fluid in a way motivated by the Cahn-Hilliard
approach to non-equilibrium dynamics and ensure that
a thermodynamically consistent state is reached
in equilibrium. The non-equilibrium dynamics is investigated
numerically and found to agree with simple analytic
predictions in both the one-phase and the two-phase region
of the phase diagram.
\end{abstract}
\vskip 0.5 truein
PACS numbers: 02.70Ns, 64.90.+b, 47.11.+j

\newpage

Since the introduction of automaton based models capable
of simulating hydrodynamic phenomena\cite{FHP}, 
lattice gas and lattice
Boltzmann techniques have proved to be extremely useful
numerical tools for investigating fluid dynamics. These
approaches have been applied to a variety of different physical
problems ranging from the simple flow of one component-fluids
to multi-component fluid dynamics in random porous media \cite{GUN},
a problem of great interest to the oil industry \cite{SAH}. 

A feature common to all lattice Boltzmann schemes is to
describe the fluid or fluid mixture by a set of distribution
functions. These probabilities are assumed to obey a 
Boltzmann equation, discrete in both time and space, which
is readily evolved numerically. The macroscopic fluid variables,
for example the density and velocity, are defined via moments
of the distribution functions, and, provided the Boltzmann
collision operator is chosen in a suitable way, 
can be shown to obey
hydrodynamic transport equations\cite{BEN}.

For a single-component fluid with an ideal equation of state,
the collision operator need only respect local conservation
of mass and momentum. This is analogous to the Boltzmann
approximation in continuum kinetic theory\cite{KT}. 
However, non-ideal fluids and 
immiscible fluid mixtures are examples of interacting systems
and to model the consequences of the interactions, 
most notably phase separation, effective collision processes
which mimic the particle interactions must be defined.
There have been several attempts to do this either by introducing
effective interactions directly between the probability
variables\cite{AZ90} or by a phenomenological rewriting
of the collision rules to favour interface formation\cite{KEL}.
Such approaches have proved successful
in a variety of applications\cite{ALEX},
but all have the disadvantage that 
the resulting macroscopic behaviour cannot,
in general, be related directly to a thermodynamic description
of the fluid\cite{SC}.

Therefore in this letter we introduce a lattice Boltzmann scheme
for simulating immiscible binary liquid mixtures which retains
the 
numerical efficiency of earlier lattice Boltzmann models of
multi-phase systems, while ensuring the equilibrium
macroscopic behaviour is
consistent with a thermodynamic description of the fluid mixture
\cite{SOY}.
Our approach was motivated by considering the 
lattice Boltzmann technique as a
mesoscopic rather than microscopic description of a fluid
which allowed us to input the conventional coarse-grained
description of non-equilibrium dynamics\cite{DOMB8}.

The basic physical variables we shall choose to work with are 
(1) the total
fluid density $\rho = \rho_1 + \rho_2$, (2) the mean fluid velocity
$\ve{u}$ and (3) the density difference between the two components
$\Delta \rho = \rho_1 - \rho_2$, where $\rho_1$ and $\rho_2$
are the individual component densities.
We introduce two distribution functions  each of which evolves 
according 
to a single relaxation time Boltzmann equation\cite{BGK}
\be
f_{i}(\ve{x}+\ve{e}_{i}\Delta t,t+\Delta t) - f_{i}(\ve{x},t)=
-\frac{1}{\tau_\rho} ( f_{i}- f_{i}^{eq}),
\ee
\be
\Delta_{i}(\ve{x}+\ve{e}_{i}\Delta t,t+\Delta t) - \Delta_{i}(\ve{x},t)=
-\frac{1}{\tau_\Delta} ( \Delta_{i}- \Delta_{i}^{eq}),
\ee
where $\tau_\rho $ and $\tau_\Delta $ are independent relaxation
parameters and
$\ve{e}_i$ is a lattice vector. 
The distribution functions are related to the physical
variables by
\be
\rho=\sum_i f_i, \ \ \ \ \ \ 
\rho\ve{u}=\sum_i f_i\ve{e}_i,\ \ \ \ \ \ 
\Delta \rho = \sum_i \Delta_i.
\ee
These three quantities are locally conserved in any collision
process and this requirement imposes three constraints on the equilibrium
distribution functions in eqns(1) and (2), namely,
\be
\sum_i f_i^{eq} = \rho,\ \ \ \ \ \  
\sum_i f_i^{eq}\ve{e}_i= \rho\ve{u},\ \ \ \ \ \ 
\sum_i \Delta_i^{eq} = \Delta\rho.
\ee
The higher moments of $f_i^{eq}$ and $\Delta_i^{eq}$
are defined so that the resulting continuum equations
take the form pertinent to a binary liquid mixture\cite{DOMB8}. 
We require that
$\Delta\rho$ obeys the Cahn-Hilliard equation supplemented by an
advection term, while the fluid as a whole obeys the Navier-Stokes 
equations with a non-ideal pressure tensor. To this end we define
\be
\sum_i \Delta_i^{eq} e_{i\alpha}= \Delta\rho u_\alpha,
\ \ \ \ \ \ \ \ 
\sum_i \Delta_i^{eq} e_{i\alpha}e_{i\beta}
= \Gamma \Delta \mu \delta_{\alpha \beta} + \Delta\rho u_\alpha u_\beta,
\ee
for the second and third moment of $\Delta_i^{eq}$, and 
\be
\sum_i f_i^{eq} e_{i\alpha}e_{i\beta} =
P_{\alpha \beta} + \rho u_\alpha u_\beta
\ee
for the third moment of $f_i^{eq}$.
In the above equations
$\Delta \mu$ is the chemical potential
difference between the two components,
$\Gamma$ is the mobility and $P_{\alpha \beta}$
is the pressure tensor. 

With these definitions of $f_i^{eq}$ and $\Delta_i^{eq}$ the 
continuum equations which follow
from expanding eqns(1) and (2) to $O({\Delta t}^2)$, are
\be
\partial_t \Delta \rho + \partial_\alpha \Delta \rho u_\alpha =
\Gamma \theta \nabla^2 \Delta \mu - \theta \partial_\alpha 
\frac{\Delta \rho}{\rho} \partial_\beta P_{\alpha \beta},
\ee
\be
\partial_t \rho +\partial_\alpha \rho u_\alpha = 0, 
\ee
\be
\partial_t \rho u_\alpha + \partial_\beta \rho u_\alpha u_\beta =
-\partial_\beta P_{\alpha \beta}
+\nu \nabla^2 \rho u_\alpha
+4\nu \partial_\alpha ( \lambda \ve{\nabla} . \rho \ve{u}),
\ee
where
\begin{eqnarray}
\theta=\Delta t c^2 \left(\tau_\Delta - \frac{1}{2}\right),& \ \ \ \ 
\nu=\Delta t c^2 \frac{1}{4}\left(\tau_\rho - \frac{1}{4}\right),& \ \ \ \
\lambda = \left(\frac{1}{2} -\frac{{\rm d}p_0}{{\rm d}\rho}\right).
\end{eqnarray}
Note that the viscosity is controlled by $\tau_\rho$ while
the introduction of the coefficient $\Gamma$ allows the
diffusion constant to be varied independently
of $\tau_\Delta$. 

If $\rho$ is constant eqns(7-9)
are Galilean invariant. However, there will be correction terms
of $O(\Delta t^3)$, and higher, which break this invariance. By
expanding eqn(2) to third order in $\Delta t$ we find a $\tau_\Delta$
dependent pre-factor $P^{(3)}(\tau_\Delta)=
\tau_\Delta^2-\tau_\Delta+\frac{1}{6}$. Thus,
by choosing $\tau_\Delta$ to be a zero of $P^{(3)}(\tau_\Delta)$ we can
eliminate all third order corrections to the diffusion equation
while retaining control over the diffusion constant
through $\Gamma$. Numerically, we find this to be an important 
requirement in order to preserve Galilean invariance as 
closely as possible.

The thermodynamic aspects of the model enter through $\Delta\mu$ and
$P_{\alpha \beta}$. Following the Cahn-Hilliard description of
non-equilibrium dynamics,
we calculate both of
these functions from the {\it equilibrium} free energy of the
fluid mixture. We choose the simplest model of a binary liquid:
two ideal gases with a repulsive interaction energy.
In terms of our model variables, the free energy functional of
the mixture takes the form
\be
\Psi = \int {\rm d^2 r} \{ \psi(\Delta\rho,\rho,T) +
\frac{\kappa}{2}(\nabla \rho)^2 +\frac{\kappa}{2} (\nabla \Delta \rho)^2
\},
\ee
in which $\psi(\Delta\rho,\rho,T)$ is the bulk free energy density at
a temperature $T$ and the second two terms give the free energy
contribution from density gradients\cite{RW}. For two interacting ideal gases
\be
\psi=
\frac{\lambda}{4}\rho (1-
\frac{\Delta \rho^2}{\rho^2}) -T\rho
+\frac{T}{2}(\rho+\Delta \rho)\log {(\frac{\rho+\Delta \rho}{2})}
+\frac{T}{2}(\rho-\Delta \rho)\log {(\frac{\rho-\Delta \rho}{2})},
\ee
where $\lambda$ measures the strength of the interaction. For
$T < T_c = \frac{1}{2} \lambda $ the bulk system phase separates into
one of two phases, symmetric in $\Delta \rho$. From this free energy
the chemical potential difference and pressure tensor follow in the
usual way\cite{RW},

\be
\Delta \mu (\Delta \rho,\rho,T)  =  -\lambda \frac{\Delta \rho}{\rho} +
T \log { \left ( \frac {1+\Delta \rho/\rho}{1-\Delta \rho/\rho}\right)}
-\kappa \nabla^2 (\Delta \rho),
\ee
\be
P_{\alpha \beta}(\vec{r}) = p(\vec{r}) \delta_{\alpha \beta} +
\kappa \frac{\partial \rho}{\partial x_{\alpha}}
\frac{\partial \rho}{\partial x_{\beta}}
+\kappa \frac{\partial \Delta \rho}{\partial x_{\alpha}}
\frac{\partial \Delta \rho}{\partial x_{\beta}},
\ee
with
\be
p(\vec{r})= \rho T -
\frac{\kappa}{2} \left( \rho \nabla^2 \rho +
\Delta \rho \nabla^2 \Delta \rho \right)
-\frac{\kappa}{2}\left (|\nabla \rho|^2
+ |\nabla \Delta \rho|^2 \right).
\ee

The distribution functions
$f_i^{eq}$ and $\Delta_i^{eq}$ must be chosen
to satisfiy eqns(4-6), which incorporate 
the thermodynamic description of the fluid. As only the
first three moments of $f_i^{eq}$ and $\Delta_i^{eq}$ are
specified, it is sufficient to expand these functions to second order
in $\ve{u}$. The coefficients in the expansion are allowed to be
not only functions of $\rho$ and $\Delta \rho$ but also
functions of their derivatives. This ensures that
$\Delta \mu$ and $P_{\alpha\beta}$ 
have the correct form in inhomogenous regions
of the fluid.

We now turn to the numerical implementation of the model in order
to test the theory presented above and to establish a practical 
limit on the scheme. Most of the simulations were performed 
on a triangular lattice with
$128 \times 128$ sites. The strength of the interaction
$\lambda$ was set to $1.1$ giving a critical temperature
$T_c=0.55$. We
chose $\kappa=0.1$, $c=1$ and worked with units in which $\Delta t =1$.
The behaviour of the system is controlled by the temperature
parameter $T$ and we have investigated the equilibrium and dynamical
properties in three distinct temperature regimes: $T>T_c, T=T_c$ and
$T<T_c$.

Above the critical temperature the equilibrium configuration is
a homogeneous mixture of the two components, characterised by
$\rho={\it constant}$ and $\Delta \rho =0$. For small variations
in $\Delta \rho$, eqn(7) can be linearised about $\Delta \rho =0$
resulting in a convective-diffusion equation with diffusion 
coefficient $D=2\theta \Gamma (T-T_c)$. To test this prediction
we have measured $D$ as a function of $\Gamma$
by monitoring the decay of a sinusoidal 
perturbation in $\Delta \rho$. 
Figure 1 shows the measured
values of $D$ as a function of $\Gamma$ for different values
of $T>T_c$. Agreement with the predicted dependence is excellent.
We have also investigated the Galilean invariance 
inherent in the model
by imposing a uniform velocity onto the initial perturbation.
We find the measured values of $D$ agree with the static case
for a wide range of velocities: D varies by less than $\frac{1}{2} \%$
for $0 < u < 0.5$. Note that our approach in the miscible regime
is an extension of the work by Flekkoy \cite{FL} to include additional
terms in $\Delta_i^{eq}$ which restore Galilean invariance.

Strictly at the critical temperature, the diffusion coefficient
discussed above is zero. We would thus expect a difference in
the dynamics at $T_c$ even though the equilibrium state is uniform.
This difference can be investigated by monitoring the decay of
the non-equilibrium surface tension, defined for a single
interface by
$\sigma = \int \left(\frac{\partial \Delta \rho}{\partial x}
\right)^2 {\rm d}x$,
where the integral is taken perpendicular to the 
gradient in $\Delta \rho$. 
Two distinct power laws
are observed for the decay of $\sigma$, depending on whether the
final temperature $T_f$ is above or equal to $T_c$: 
$\sigma \sim t^{-\frac{1}{2}}$ for $T_f>T_c$ and
$\sigma \sim t^{-\frac{1}{4}}$ for $T_f=T_c$. This is in agreement
with the predictions of
Ma {\it et. al.}\cite{MA}
and indicates that the system is in the 
Model B universality class, in a regime where hydrodynamics
is unimportant. Earlier lattice Boltzmann models of binary
mixtures were unable to reproduce such details.

Finally, below $T_c$ the fluid phase separates into two
distinct phases, symmetric about $\Delta \rho = 0$.
Figure 2 shows typical interfacial profiles
between two phase characterised
by $\Delta \rho > 0$ and $\Delta \rho < 0$. The width of the interface
can be tuned by varying $T$ (or the interfacial energy $\kappa$) and
chosen so that lattice effects are unimportant. We also show
the variation of the total density $\rho$ in the interfacial
region; typically, this variation is less then 2\%. Far from the
interfacial region, the uniform values of $\Delta \rho$ define the
coexisting phases. 
The coexistence curve as a function of $T$ is
shown in figure 3. The observed values of $\Delta \rho$
are seen to be in agreement with the theoretical prediction
derived by minimising the bulk free energy, eqn(12).

We have verified that deviations from Galilean invariance are very small
even in the immiscible regime: there is little change
to either the coexisting values of $\Delta \rho$ or the structure
of the interface if a uniform velocity is imposed on the system.
The velocity dependence of the coexistence curve is also show
in figure 3.
A well know problem common to all multi-phase simulation
techniques is the generation of spurious velocity currents in the
interfacial region.
We have measured this velocity field within our model 
and find its magnitude appears to be substantially lower than the 
values that have been observed
using other schemes\cite{KEL,SOY}. For example, for
the interfaces shown in figure 2, $|\ve{u}|$ is 
$O(10^{-10})$.
Lastly, we have checked that Laplace's Law holds for spherical
domains, a feature which is ensured 
in the continuum by our choice of pressure
tensor. A detailed study of the collective dynamics during
spinodal decomposition will be presented elsewhere\cite{WIL}.

To summarise: we have devised and tested a lattice Boltzmann scheme
for simulating
a binary liquid mixture which 
leads to a thermodynamically
consistent equilibrium state.
Three distinct dynamical regimes are
observed; above, at and below the critical temperature, 
and the agreement between our simulation
results and theoretical
predictions is excellent for a wide range of system parameters.
Furthermore, the effects of correction terms 
which break Galilean invariance
are seen to be small, even in the presence of sharp interfaces.

We are indebted to William Osborn and Jayanth Banavar for numerous
valuable discussions. EO acknowledges support from the European
community and MRS and JMY from the E. P. S. R. C.

\newpage

FIGURE CAPTIONS

\begin{description}

\item[Figure 1] The diffusion coefficient $D$ as a function of the
parameter $\Gamma$ for three different values of the temperature.
The points represent results from simulations:
open circles T=0.8, filled square T=0.7 and filled circles T=0.6.
The solid lines show the analytic value
$D= 2\Gamma \theta \left (T-T_c\right)$.

\item[Figure 2] Profiles normal to a flat interface
of the binary mixture for three different values
of the temperature: T=0.498 (solid), 
T=0.511 (dotted) and T=0.526
(dashed). The parameter $\kappa$ is fixed to the value $0.1$.
Note that the variation of $\rho$ in the interfacial region is
small; the `spikes' 
around $x=0$ are an artifact of lattice discretisation. 

\item[Figure 3] Coexistence curve as a function of the temperature $T$
for $\lambda = 1.1$. The different points denote simulations at different
uniform fluid velocity: filled squares u=0.0, open triangles u=0.1 and 
open hexagons u=0.2.
The solid line is the analytic curve obtained by minimising the bulk
free energy.

\end{description}

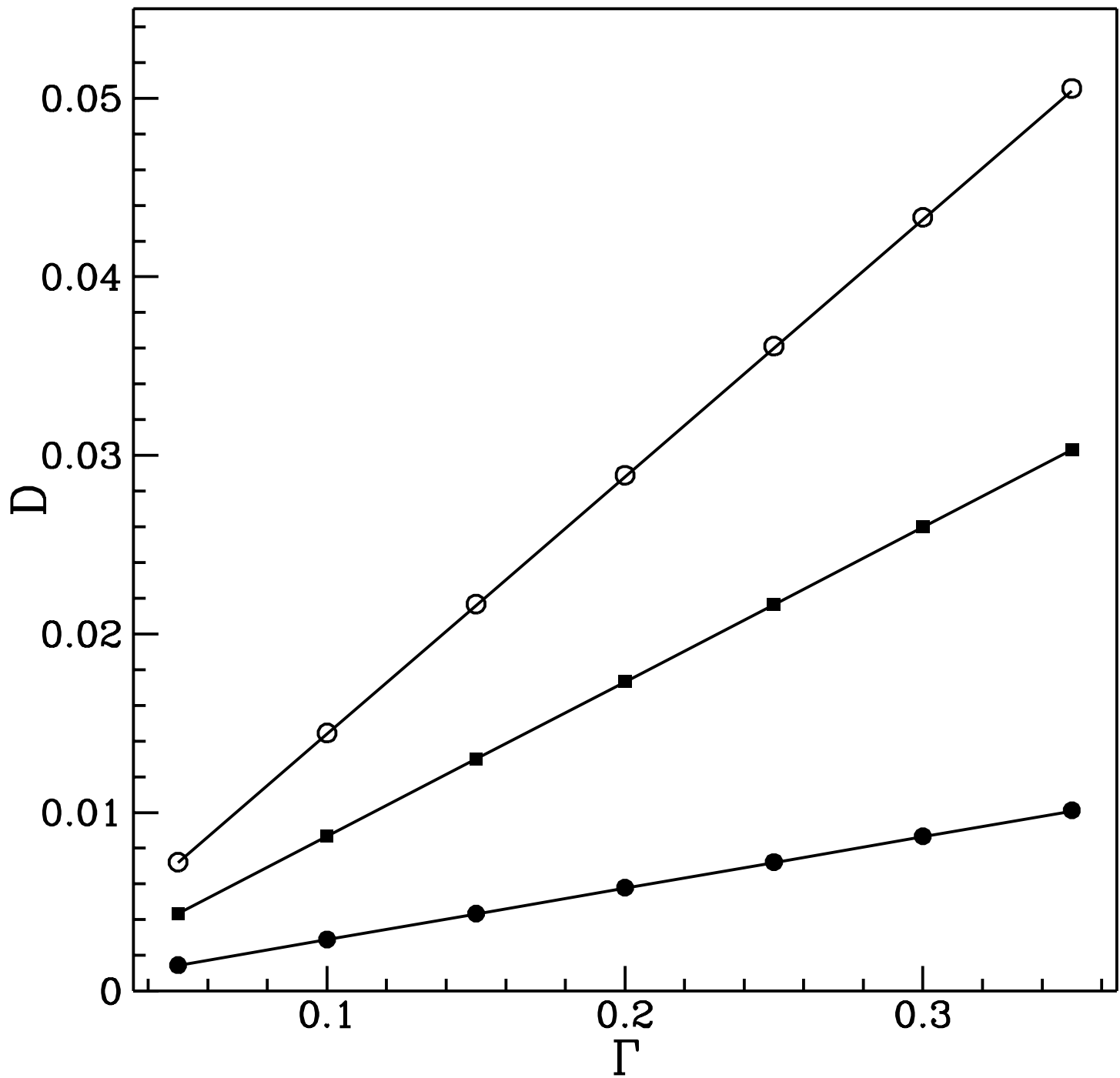
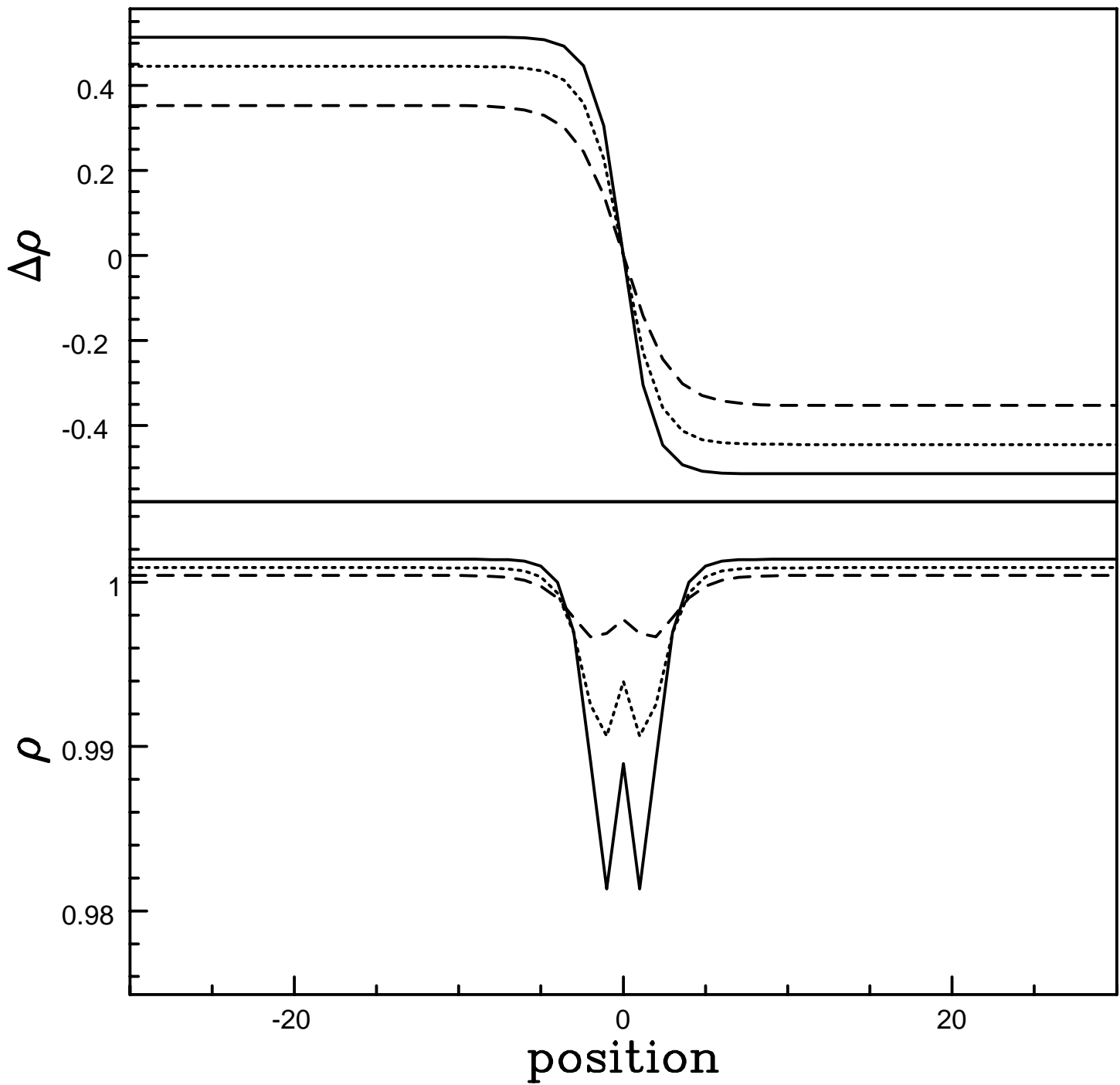
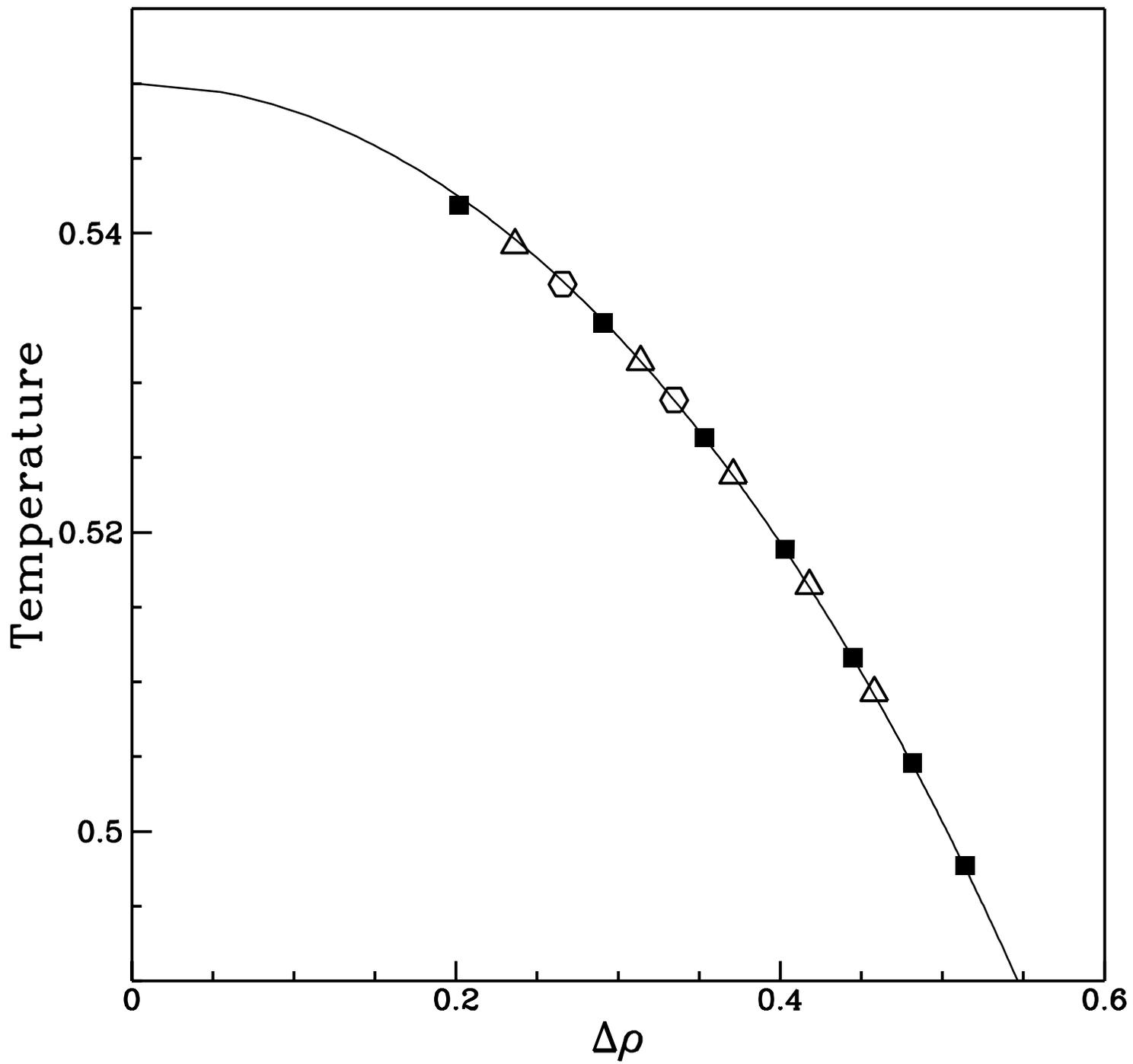
\end{document}